# Noctalgia (sky grief): Our Brightening Night Skies and Loss of Environment for Astronomy and Sky Traditions

**Published as a [*Science* e-Letter in response to the Introduction](#) of the [Special Section on light pollution in the June 16, 2023 issue of *Science*](#)**

Aparna Venkatesan
Professor, Department of Physics and Astronomy,
and co-Director of the Tracy Seeley Center for Teaching Excellence,
University of San Francisco, San Francisco, CA

John C. Barentine
Dark Sky Consulting, LLC, Tucson, AZ

Fifty years after the first mention of light pollution in *Science* (*1*), the journal recently elevated this topic to its cover (16 June 2023), highlighting the large impact on human and ecological health, circadian rhythms, migratory patterns, and more. Our diminishing ability to view the nighttime sky due to rapidly rising human-made light pollution is part of the palpable keening of all that is passing each day, including the still-increasing global loss of life to COVID-19 and the loss of environment arising from ever-accelerating climate change.

The author Paul Bogard has written eloquently not only of the loss of dark night skies but also [of our daytime skies](#) from the smoke and ash of increasingly catastrophic wildfires, a global phenomenon impacting much of humanity in summer 2023. Bogard recently edited *Solastalgia* (*2*), a collection of essays that attempts to capture the growing undercurrent of ineffable grief for our loss of individual "home environments": habitats, resources, homelands, and now the skies themselves. We offer here the term *noctalgia* to express "sky grief" for the accelerating loss of the home environment of our shared skies, a disappearance felt globally and deserving its own field of study of "nyctology" (*3*).

This represents far more than mere loss of environment: we are witnessing loss of heritage, place-based language, identity, storytelling, millennia-old sky traditions and our ability to conduct traditional practices grounded in the ecological integrity of what we call home.  This is especially true for those disproportionately impacted by climate change, the COVID-19 pandemic, and broader social inequities, including Indigenous communities. As most cyclical indicators involving migratory birds/animals, weather patterns, pollinators and native plants are increasingly altered with climate change, the skies effectively represent a last stand of heritage, calendaring, language, and food sovereignty for many global communities.

How can we invite collaborative solutions in the face of this unprecedented threat to the skies, when so many in 2023 are experiencing crisis fatigue and young people are losing hope for the planet's future? As some of the *Science* articles note, ground-based light pollution has several promising near-term solutions. Impacts to our skies from large satellite constellations and space debris (*4,5*) are more complicated to address given the current regulatory landscape and (lack of) legal protections for space and skies as an environment (*6*). Some of the next steps could be: expanding such protections and globally coordinated domestic and international policies for the skies; designation of the skies as intangible cultural heritage by the United Nations; and, expanding the language/protections associated with Earth jurisprudence and the 'Rights of Nature' (*7*).

The daytime *and* nighttime skies deserve protection as a globally shared heritage, one we hope to leave as a legacy for those who will come long after us. Otherwise, this Letter to *Science* may represent a reluctant yet anticipated apologia from our elegiac time to future generations of scientists, storytellers, artists and knowledge-holders.